\documentclass[3p,times,procedia]{elsarticle}
\usepackage{nupha_ecrc}
\usepackage[utf8]{inputenc}

\volume{00}

\firstpage{1}

\journalname{Nuclear Physics A}

\runauth{}

\jid{nupha}

\jnltitlelogo{Nuclear Physics A}

\usepackage{amssymb}
 \usepackage{amsthm}
 \usepackage{amsmath}

\usepackage{lineno}

\usepackage[figuresright]{rotating}

\begin{document}

\begin{frontmatter}



\dochead{XXVIIth International Conference on Ultrarelativistic Nucleus-Nucleus Collisions\\ (Quark Matter 2018)}

\title{Elliptic flow of identified hadrons in small collisional systems measured with ALICE}


\author{Vojtěch Pacík\\ on behalf of the ALICE Collaboration}

\address{Niels Bohr Institute, University of Copenhagen, Blegdamsvej 17, Denmark}

\begin{abstract}
Recent observations of long-range multi-particle azimuthal correlations in p--Pb and high multiplicity pp collisions provided new insights into collision dynamics and opened a possibility to study collective effects in these small systems.
New measurement of $p_\mathrm{T}$-differential elliptic flow ($v_{2}$) coefficient for inclusive charged hadrons as well as a variety of identified particle species in p--Pb collisions at $\sqrt{s_\mathrm{NN}}=5.02$~TeV recorded by ALICE during the LHC Run 2 data taking are presented.
Besides high precision measurements of $v_2(p_\mathrm{T})$ for $\mathrm{\pi^\pm,~K^\pm}$ and $\mathrm{p(\bar{p})}$, the very first results for $\mathrm{K_{S}^{0},~\Lambda(\bar{\Lambda})}$ and $\phi$ are shown.
In order to eliminate non-flow contamination, a pseudorapidity separation between correlated particles is applied as well as subtraction of remaining non-flow estimate based on a measurement of minimum-bias pp collisions at $\sqrt{s}=13$~TeV.
Moreover, reported characteristic mass ordering and approximate NCQ and $\mathrm{KE_T}$ scaling of $v_{2}$ allows to test various theoretical models, constrain the initial conditions, and probe the origin of collective behavior in small collision systems.
\end{abstract}


\end{frontmatter}


\section{Introduction}
\label{Sec:Introduction}

Formation of Quark-Gluon plasma (QGP), strongly-interacting nuclear matter created under extreme conditions such as collisions of heavy ions at the Large Hadron Collider (LHC), is well established and supported by plethora of measurements. In such collisions, initial spacial anisotropy evolves into momentum anisotropy by hydrodynamical expansion of the created medium. The azimuthal momentum distribution of emitted particles can be decomposed into a Fourier expansion \eqref{eq:Fourier} with anisotropic flow coefficients $v_n = \langle \cos{n(\varphi - \Psi_{n})} \rangle$, where $\varphi$ is azimuthal angle and $\Psi_{n}$ is symmetry plane of $n$-th harmonic~\cite{Voloshin:1994mz}.
\begin{equation}\label{eq:Fourier}
E\frac{\mathrm{d}N}{\mathrm{d}\vec{p}} = \frac{1}{2\pi}\frac{\mathrm{d}^2N}{p_\mathrm{T}\mathrm{d}p_\mathrm{T}\mathrm{d}y}\Bigg[1+2 \sum_{n=1}^{\infty}{v_{n} \cos{[n(\varphi-\Psi_{n})]}}\Bigg]
\end{equation}
Specifically, results of second order coefficient $v_2$, referred to as elliptic flow, are presented in this contribution.

Recent measurements of anisotropic flow coefficients in high multiplicity p--Pb and pp collisions at the LHC exhibit similar features as ones observed in case of heavy-ion collisions. Observation of non-zero values of $v_n$ indicate a presence of collective behavior in small collision systems, even though QGP creation was not expected in such a limited volume. Measurements of $p_\mathrm{T}$-differential $v_n$ coefficients are used to investigate the origin of such collectivity, which remains an open question.
Moreover, study of this observable for various particle species provides a unique tool to study the effect of collective phenomena in the context of particle production mechanisms.

\section{Analysis details}
\label{Sec:Analysis}
The data sample used for this analysis contains 600 million minimum-bias p--Pb collisions at $\sqrt{s_\mathrm{NN}}=5.02$~TeV and 166 million minimum-bias pp collisions at $\sqrt{s}=13$~TeV recorded by the ALICE detector~\cite{Aamodt:2008zz} in 2016 during Run 2 phase of the LHC operation.
The sample is divided into multiple event classes based on the multiplicity distribution measured by the V0A detector with a pseudorapidity coverage of $2.8 < \eta < 5.1$ (Pb-going direction).

Charged particles are reconstructed using tracking information from Inner Tracking System (ITS) and Time-Projection Chamber (TPC) with a coverage of pseudorapidity $|\eta| < 0.8$ and full azimuthal angle. Particle identification (PID) of $\mathrm{\pi^{\pm},~K^\pm~and~p(\bar{p})}$ is performed based on signal from TPC and Time-of-Flight (TOF) detectors using Bayesian approach described in~\cite{Adam:2016acv}. The purity of the selected particles is above 95\% over the whole reported momentum range (80\% for $\mathrm{K^{\pm}}$ at $p_\mathrm{T} > 3.5~\mathrm{GeV}/c$). Due to their neutral charge and relatively short life-time, $\mathrm{K_{S}^{0}}$, $\Lambda(\bar{\Lambda})$ and $\phi$ cannot be detected directly. Instead, selection is performed on candidates reconstructed via their decay products on a statistical basis. Specifically, the following hadronic decays are considered: $\mathrm{K_{S}^{0}} \rightarrow \pi^+ + \pi^-$, $\mathrm{\phi \rightarrow K^+ + K^-}$ and $\mathrm{\Lambda \rightarrow \pi^- + p}$ ($\mathrm{\bar{\Lambda}\rightarrow\pi^{+}+\bar{p}}$).
 In order to suppress combinatorial background, PID of decay products and selection criteria on characteristic decay topology of $\mathrm{K_{S}^{0}}$ and $\Lambda(\bar{\Lambda})$ are applied.

The $p_\mathrm{T}$-differential $v_2$ coefficient is obtained from measurement of 2-particle Q-cumulants~\cite{Bilandzic:2010jr} using the generic framework implementation including the non-uniform acceptance correction introduced in~\cite{Bilandzic:2013kga}. Inclusive charged particles within $0.2 < p_\mathrm{T} < 3~\mathrm{GeV}/c$ range were selected as reference flow particles.

Measurements of two-particle correlations in small collisional systems are very sensitive to short-range contribution not related to the common symmetry plane, such as jets and resonance decays, commonly referred to as non-flow. To suppress such bias, the sub-event method, where pseudorapidity separation ($\Delta\eta$ gap) is introduced between the correlated particles, is used. Additionally, in order to remove remaining contamination, subtraction on cumulant level is performed using data of minimum-bias pp collisions. The subtraction is done according to
\begin{equation}\label{eq:Subtraction}
v_{2}^\mathrm{pPb,sub}(p_\mathrm{T}) = \frac{d_{2}^\mathrm{pPb}(p_\mathrm{T}) - k \cdot d_{2}^\mathrm{pp}(p_\mathrm{T})}{\sqrt{c_{2}^\mathrm{pPb} - k \cdot c_{2}^\mathrm{pp}}},
\end{equation}
 where $c_{2}$ and $d_{2}$ denote integrated (reference flow) and differential cumulants, respectively~\cite{Bilandzic:2010jr}. Prior to subtraction, non-flow estimation is scaled by a ratio of mean event multiplicities $k=\left\langle M \right\rangle^\mathrm{pp}/\left\langle M \right\rangle^\mathrm{pPb}$ based on its scaling properties~\cite{Voloshin:2008dg} to account for different system size of p--Pb and pp collision.

\section{Results}
\label{Sec:Results}
The results of non-flow subtracted $p_\mathrm{T}$-differential elliptic flow coefficients of inclusive charged hadrons $\mathrm{h}^{\pm}$ and identified  $\mathrm{\pi}^{\pm},~\mathrm{K}^{\pm},~\mathrm{K_{S}^{0}},~\mathrm{p}(\bar{\mathrm{p}}),~\phi$ and $\Lambda(\bar{\Lambda})$ with pseudorapidity separation of $|\Delta\eta| > 0.4$ in p--Pb collisions at $\sqrt{s_\mathrm{NN}} = 5.02$~TeV are presented.

\begin{figure}[!tb]
  \centering
  \begin{minipage}{0.49\textwidth}
    \centering
    \includegraphics[width=1.0\textwidth]{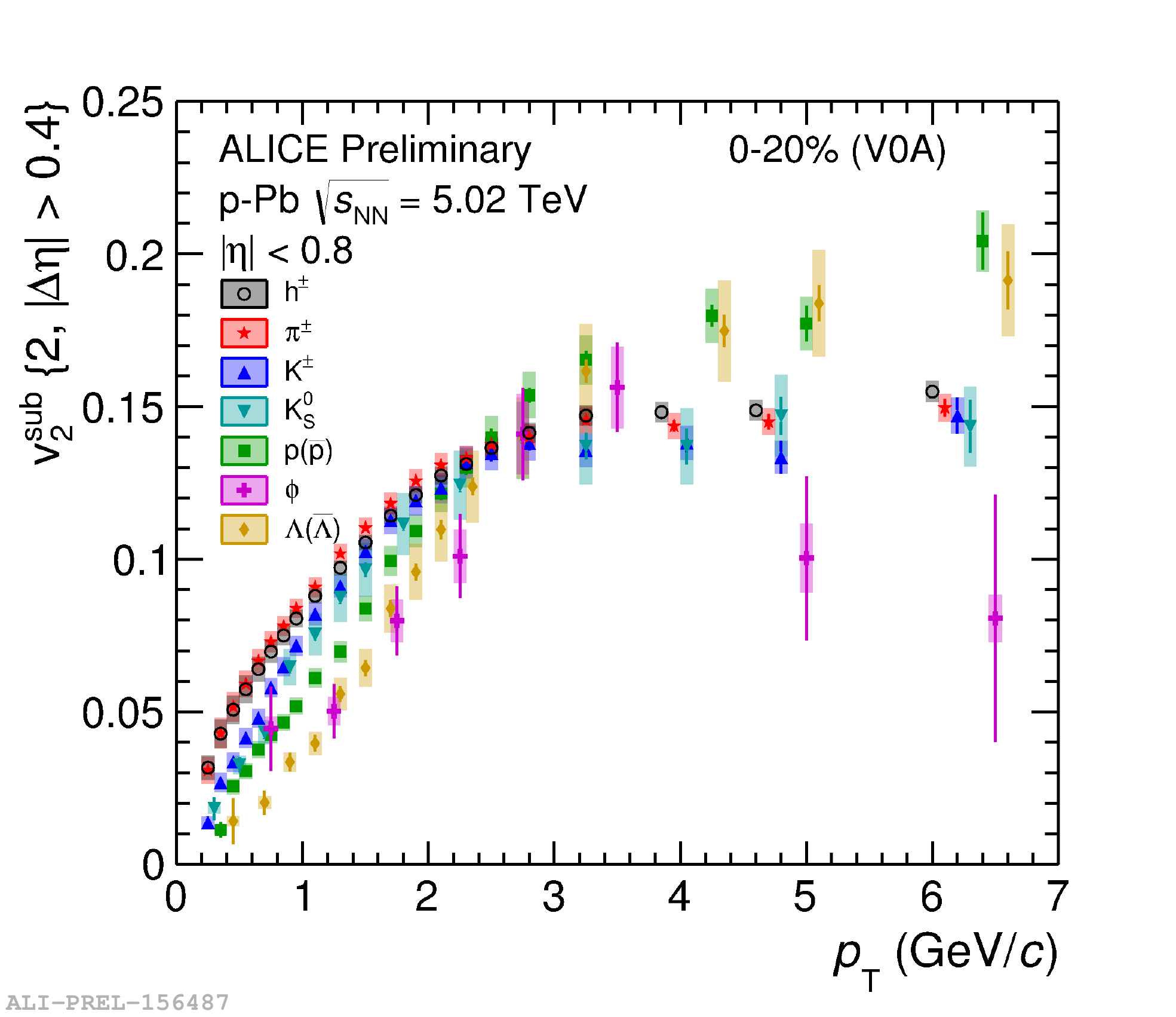}
  \end{minipage}
  \begin{minipage}{0.49\textwidth}
    \centering
    \includegraphics[width=1.0\textwidth]{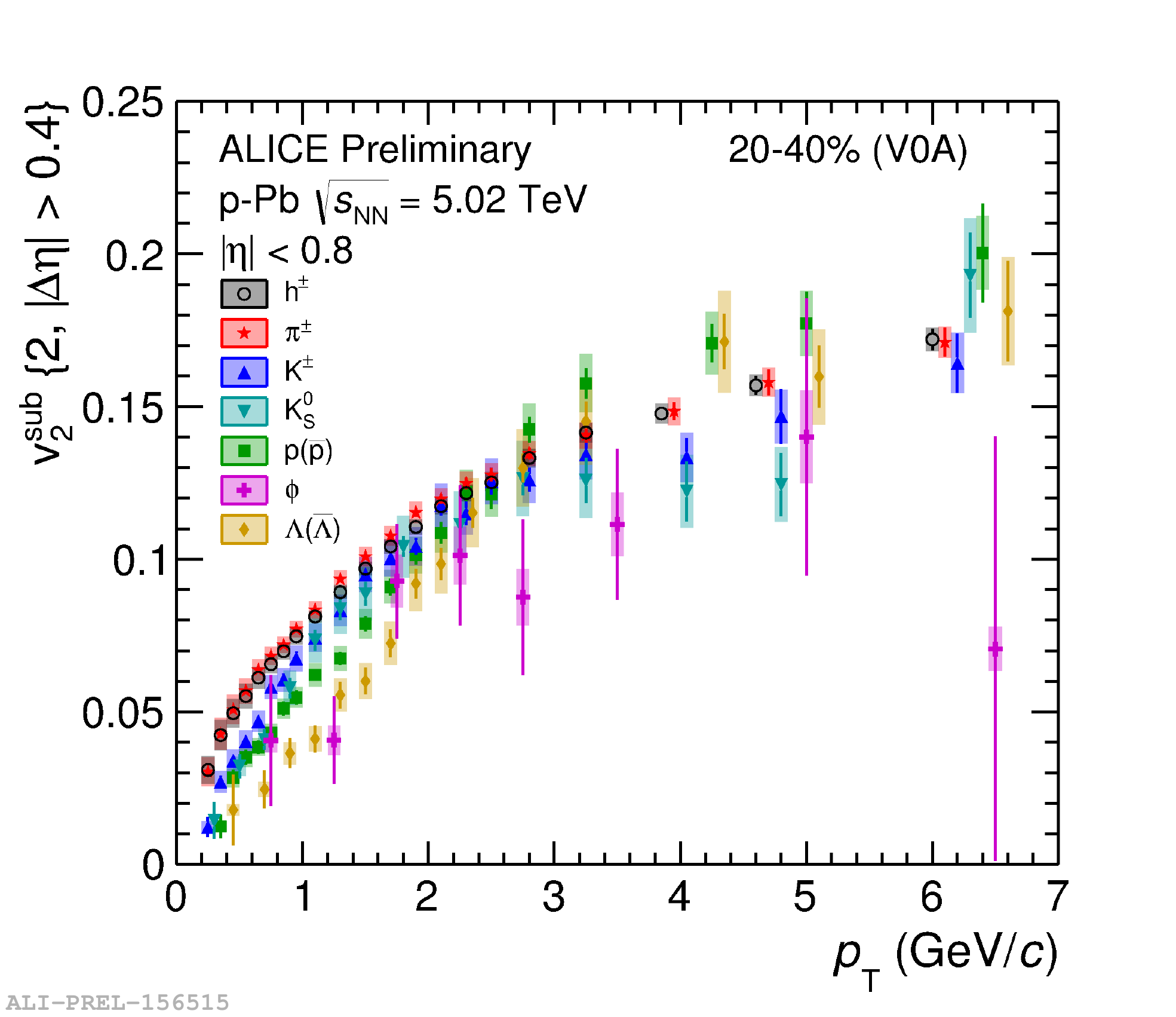}
  \end{minipage}
\caption{Results of $p_\mathrm{T}$-differential $v^\mathrm{sub}_{2}\{2,|\Delta\eta| > 0.4\}$ of $\mathrm{h}^{\pm},~\mathrm{\pi}^{\pm},~\mathrm{K}^{\pm},~\mathrm{K_{S}^{0}},~\mathrm{p}(\bar{\mathrm{p}}),~\phi$ and $\Lambda(\bar{\Lambda})$ in p--Pb collisions at $\sqrt{s_\mathrm{NN}} = 5.02$~TeV with non-flow subtraction performed with measurements of minimum-bias pp collisions at $\sqrt{s}=13$~TeV for V0A multiplicity classes: 0-20\% (left) and 20-40\% (right).}
\label{Fig:v2sub}
\end{figure}

The results for two events classes based on forward multiplicity measured with V0A detector are shown in Fig.~\ref{Fig:v2sub}: 0-20\% events with highest multiplicity (left) and 20-40\% (right). For $p_\mathrm{T} \leq 2.5~\mathrm{GeV}/c$, a clear mass ordering of $v_2(p_\mathrm{T})$ is observed, which is an indication of strong radial expansion with common velocity pushing heavier particles towards higher momenta. This trend is qualitatively consistent with the results of hydrodynamic models~\cite{Werner:2013ipa,Bozek:2013ska}. In the region of $2.5 < p_\mathrm{T} \lesssim 6~\mathrm{GeV}/c$, particles are grouped based on their constituent quark content with a clear separation between the higher $v_{2}(p_\mathrm{T})$ of baryons and the lower values for mesons. In the picture of heavy-ion collision, such splitting effect is usually attributed to parton coalescence~\cite{Molnar:2003ff} or recombination mechanism of particle production~\cite{Hwa:2002tu}.

\begin{figure}[h]
   \centering
   \begin{minipage}{0.49\textwidth}
     \centering
     \includegraphics[width=1.0\textwidth]{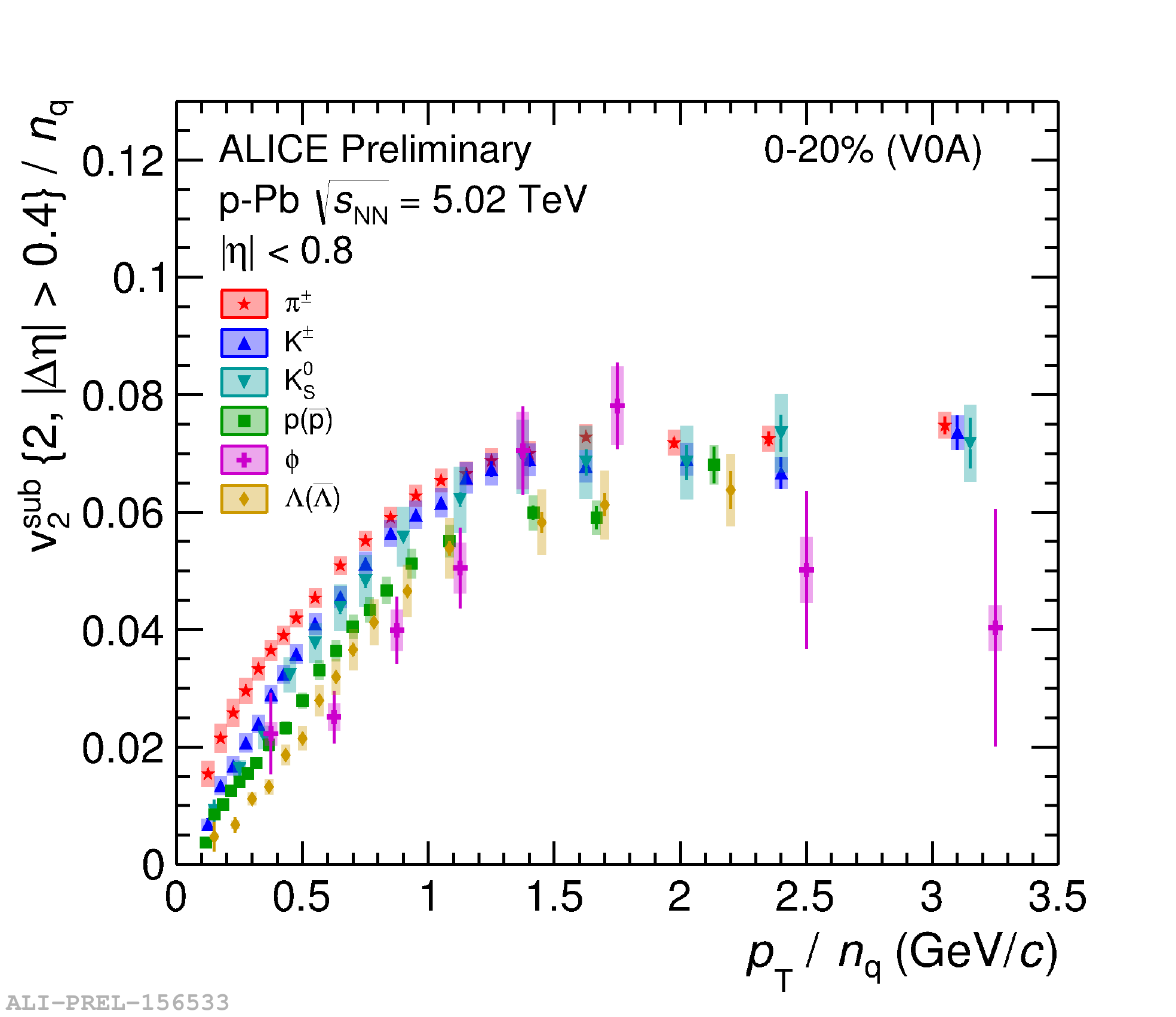}
   \end{minipage}
   \begin{minipage}{0.49\textwidth}
     \centering
     \includegraphics[width=1.0\textwidth]{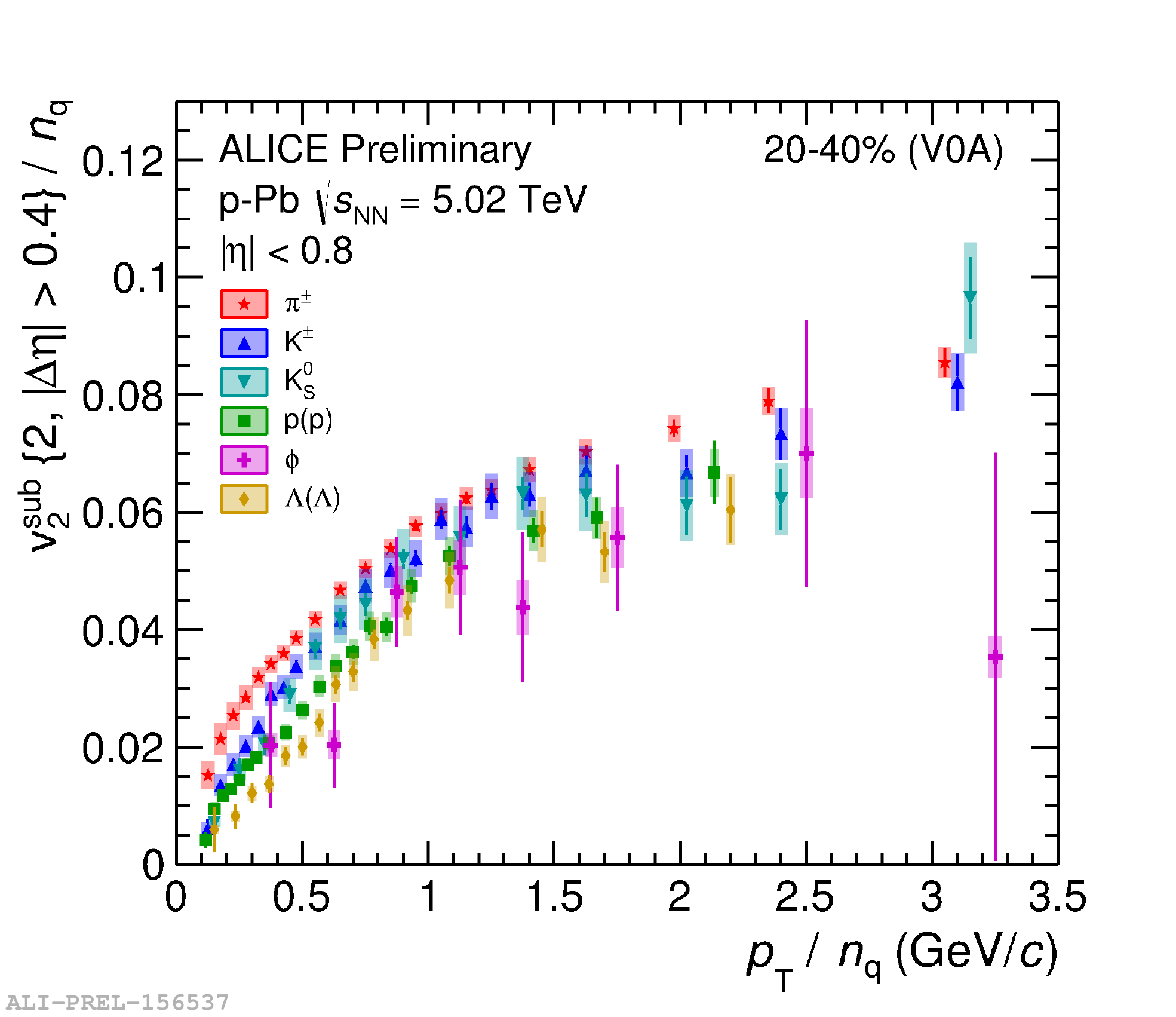}
   \end{minipage}
   \begin{minipage}{0.49\textwidth}
     \centering
     \includegraphics[width=1.0\textwidth]{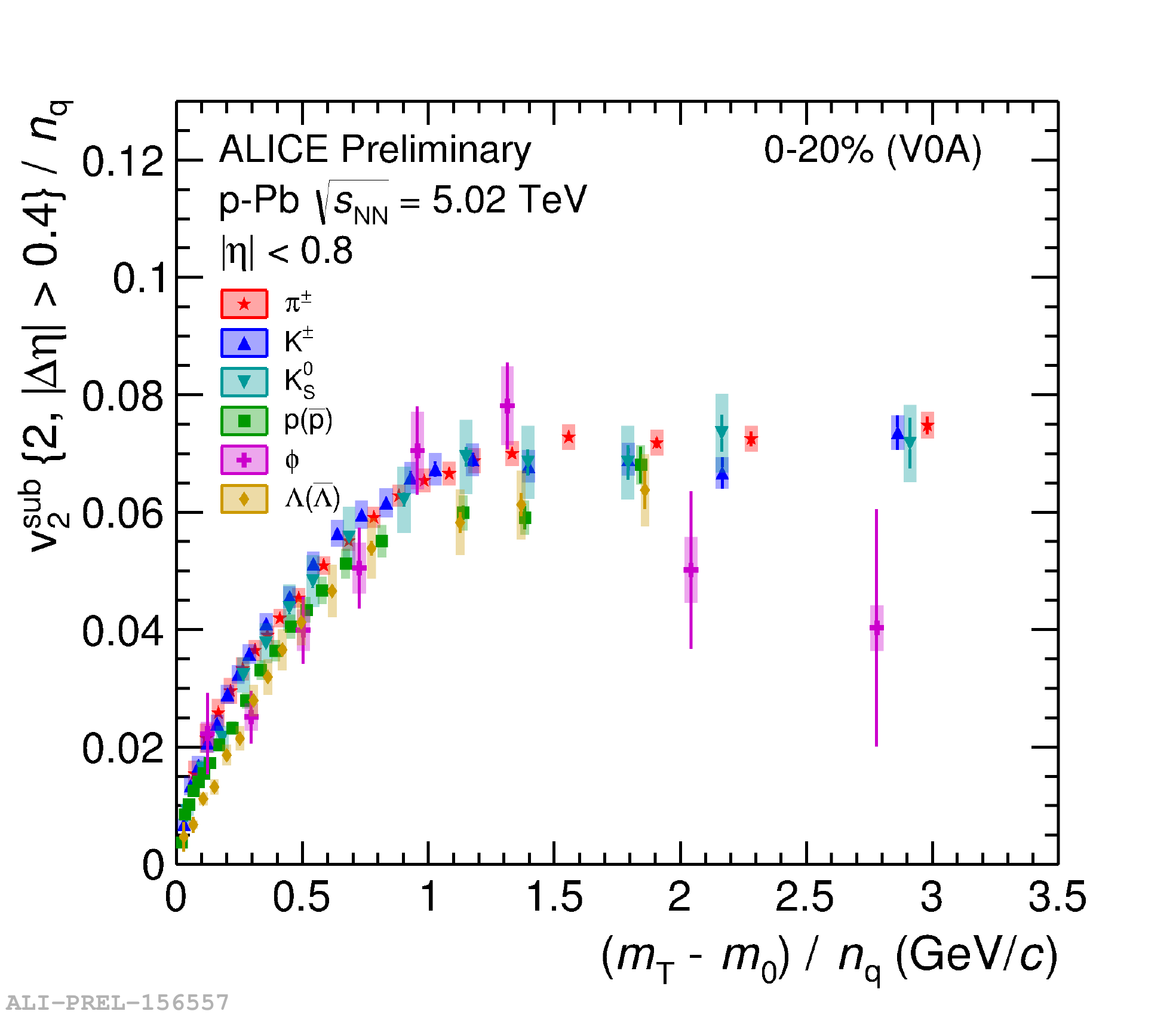}
   \end{minipage}
   \begin{minipage}{0.49\textwidth}
     \centering
     \includegraphics[width=1.0\textwidth]{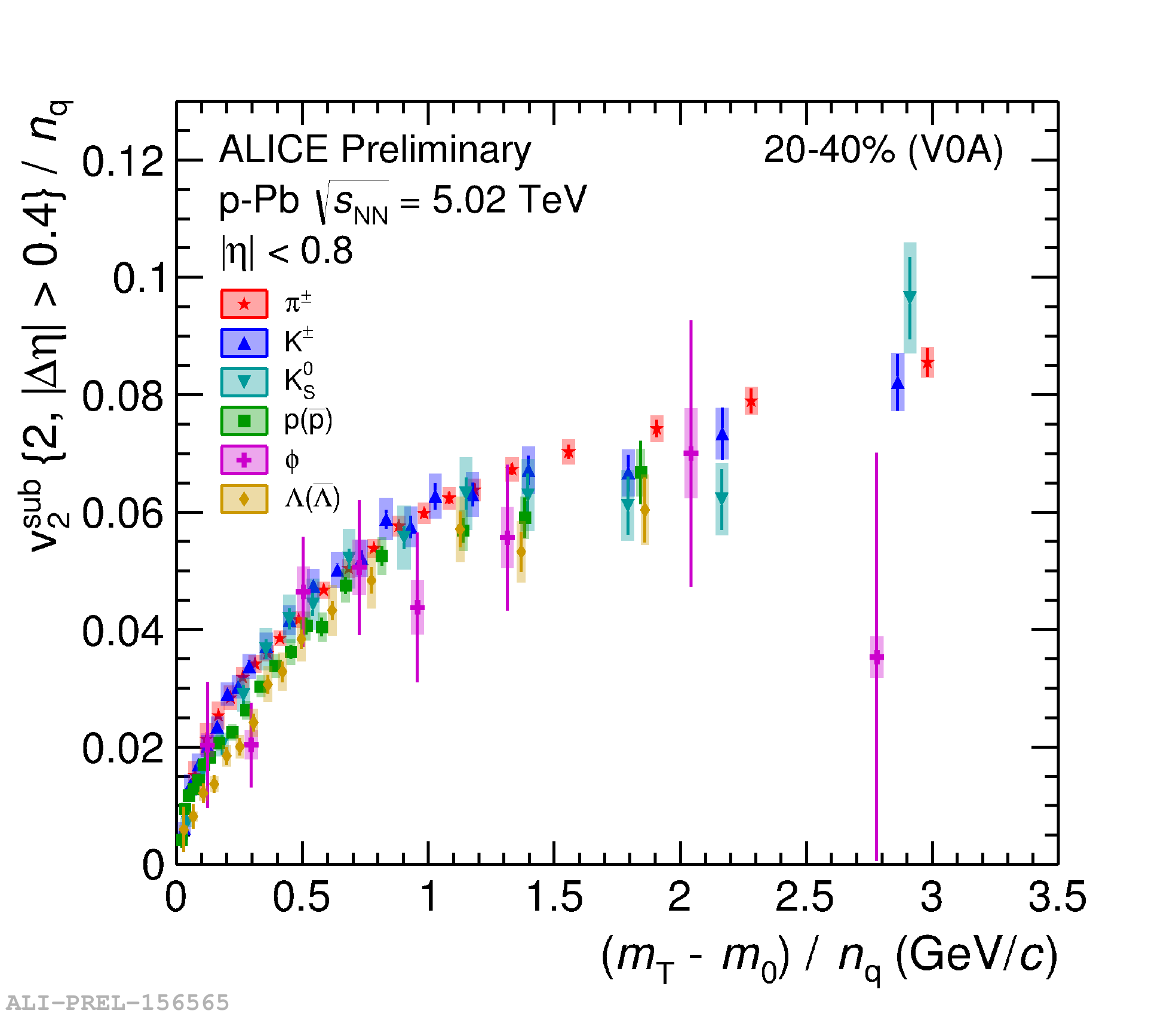}
   \end{minipage}
 \caption{Test of scaling properties of $v_{2}$ coefficients. Upper row: both $v_{2}^\mathrm{sub}$ and $p_\mathrm{T}$ are scaled by number of constituent quarks $n_\mathrm{q}$ (NCQ scaling). Bottom row: transverse kinetic energy ($\mathrm{KE_T}$) scaling, where rest mass $m_\mathrm{0}$ of individual species is subtracted from $m_\mathrm{T}$.}
 \label{Fig:ncq}
 \end{figure}

To test the extent of scaling properties of the elliptic flow coefficient originally reported by RHIC experiments~\cite{Adler:2003kt}, both $v_{2}$ and $p_\mathrm{T}$ were divided by the number of constituent quarks $n_\mathrm{q}$ (NCQ scaling). The result of such scaling is shown in the upper row of Fig.~\ref{Fig:ncq}. Moreover, to extend the scaling to low $p_\mathrm{T}$ region and to account for the observed mass hierarchy of $v_2$, transverse kinetic energy
\begin{equation} \label{eq:KEt}
\mathrm{KE_T} = m_\mathrm{T} - m_\mathrm{0} = \sqrt{p_\mathrm{T}^2+m_\mathrm{0}^2} - m_\mathrm{0}
\end{equation}
is used instead of $p_\mathrm{T}$ ($\mathrm{KE_T}$ scaling) and the results are reported in the bottom row of~Fig.~\ref{Fig:ncq}.
In both cases, all particle species show a universal trend indicating a presence of collective behavior on the partonic level in p--Pb collisions, even though the observed scaling is only approximate.

\section{Conclusion}
\label{Sec:Conclusion}
New measurements of $v_{2}(p_\mathrm{T})$ using 2-particle cumulants of both inclusive and identified charged hadrons in p--Pb collisions at $\sqrt{s_\mathrm{NN}} = 5.02$~TeV using the LHC Run 2 data are presented.

The reported measurements of elliptic flow coefficients in p--Pb collisions exhibit similar features as previously reported results from Pb--Pb collisions~\cite{Abelev:2014pua, Acharya:2018zuq}. Specifically, they confirm the mass ordering in the low $p_\mathrm{T}$ region followed by the baryon/meson grouping at intermediate $2.5 < p_\mathrm{T} \lesssim 6~\mathrm{GeV}/c$ together with only approximate NCQ and $\mathrm{KE_{T}}$ scaling. This further supports that collective phenomena are present in high multiplicity collisions of small systems. However, whether they are manifestations of initial or final state effects is not yet fully understood.
The unprecedented precision of reported $v_2(p_\mathrm{T})$ results presents an invaluable tool to further constrain theoretical models, test their validity and penultimately will help to disentangle the origin of such collectivity.





\bibliographystyle{elsarticle-num}
\bibliography{proceedings}

\begin{thebibliography}{10}
\expandafter\ifx\csname url\endcsname\relax
  \def\url#1{\texttt{#1}}\fi
\expandafter\ifx\csname urlprefix\endcsname\relax\def\urlprefix{URL }\fi
\expandafter\ifx\csname href\endcsname\relax
  \def\href#1#2{#2} \def\path#1{#1}\fi

\bibitem{Voloshin:1994mz}
S.~Voloshin, Y.~Zhang, {Flow study in relativistic nuclear collisions by
  Fourier expansion of Azimuthal particle distributions}, Phys. C70 (1996)
  665--672.

\bibitem{Aamodt:2008zz}
K.~Aamodt, et~al., {The ALICE experiment at the CERN LHC}, JINST 3 (2008)
  S08002.

\bibitem{Adam:2016acv}
J.~Adam, et~al., {Particle identification in ALICE: a Bayesian approach}, Eur.
  Phys. J. Plus 131~(5) (2016) 168.

\bibitem{Bilandzic:2010jr}
A.~Bilandzic, R.~Snellings, S.~Voloshin, {Flow analysis with cumulants: Direct
  calculations}, Phys. Rev. C83 (2011) 044913.

\bibitem{Bilandzic:2013kga}
A.~Bilandzic, C.~H. Christensen, K.~Gulbrandsen, A.~Hansen, Y.~Zhou, {Generic
  framework for anisotropic flow analyses with multiparticle azimuthal
  correlations}, Phys. Rev. C89~(6) (2014) 064904.

\bibitem{Voloshin:2008dg}
S.~A. Voloshin, A.~M. Poskanzer, R.~Snellings, {Collective phenomena in
  non-central nuclear collisions}, Landolt-Bornstein 23 (2010) 293--333.

\bibitem{Werner:2013ipa}
K.~Werner, M.~Bleicher, B.~Guiot, I.~Karpenko, T.~Pierog, {Evidence for Flow
  from Hydrodynamic Simulations of $p$-Pb Collisions at 5.02 TeV from $\nu_2$
  Mass Splitting}, Phys. Rev. Lett. 112~(23) (2014) 232301.

\bibitem{Bozek:2013ska}
P.~Bozek, W.~Broniowski, G.~Torrieri, {Mass hierarchy in identified particle
  distributions in proton-lead collisions}, Phys. Rev. Lett. 111 (2013) 172303.

\bibitem{Molnar:2003ff}
D.~Molnar, S.~A. Voloshin, {Elliptic flow at large transverse momenta from
  quark coalescence}, Phys. Rev. Lett. 91 (2003) 092301.

\bibitem{Hwa:2002tu}
R.~C. Hwa, C.~B. Yang, {Scaling behavior at high p(T) and the p / pi ratio},
  Phys. Rev. C67 (2003) 034902.

\bibitem{Adler:2003kt}
S.~S. Adler, et~al., {Elliptic flow of identified hadrons in Au+Au collisions
  at s(NN)**(1/2) = 200-GeV}, Phys. Rev. Lett. 91 (2003) 182301.

\bibitem{Abelev:2014pua}
B.~B. Abelev, et~al., {Elliptic flow of identified hadrons in Pb-Pb collisions
  at $ \sqrt{s_{\mathrm{NN}}}=2.76 $ TeV}, JHEP 06 (2015) 190.

\bibitem{Acharya:2018zuq}
S.~Acharya, et~al., {Anisotropic flow of identified particles in Pb-Pb
  collisions at $\mathbf{\sqrt{{\textit s}_{\rm NN}}}=5.02$ TeV,}\href
  {http://arxiv.org/abs/1805.04390} {\path{arXiv:1805.04390}}.

\end{thebibliography}







\end{document}